\documentclass[11pt,letterpaper]{article}
\usepackage{graphicx}
\usepackage{bm}
\usepackage{amssymb}
\usepackage{color}
\usepackage{amsmath}
\usepackage[numbers]{natbib}
\usepackage{enumerate}
\usepackage{amstext}
\usepackage{footmisc}
\usepackage{latexsym}
\usepackage[usenames,dvipsnames]{xcolor}
\usepackage[colorlinks=true,citecolor=Cerulean,linkcolor=RubineRed,urlcolor=Cerulean]{hyperref}
\usepackage{multibib}
\usepackage{ fixfoot}
\usepackage{fancyhdr}
\newcommand{\pac}[1]{ \left\{ #1 \right\} }
\newcommand{\pap}[1]{\left( #1 \right)}
\newcommand{\pas}[1]{\left[#1 \right]}

\newcommand{\pabs}[1]{ \left| #1 \right| }
\newcommand{\ket}[1]{ \left| #1 \rangle\right.}
\newcommand{\bra}[1]{  \left.\langle #1  \right|}

\DeclareFixedFootnote{\rep}{Electronic address: \href{mailto:fernandojavier.gomez@iff.csic.es}{fernandojavier.gomez@iff.csic.es}}

\newcommand{\beq}{\begin{equation}}
\newcommand{\eeq}{\end{equation}}
\newcommand{\beqa}{\begin{eqnarray}}
\newcommand{\eeqa}{\end{eqnarray}}

\newcommand{\blue}[1]{{\color{blue} #1}}

\begin{document}
\title{{\bf Experimental validation of the Kibble-Zurek Mechanism on a Digital Quantum Computer}}
\author{Santiago Higuera-Quintero\href{https://orcid.org/0000-0001-6905-4662}{\includegraphics[scale=0.45]{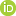}}\,$^{1}$, Ferney J. Rodr\'iguez\href{https://orcid.org/0000-0001-5383-4218}{\includegraphics[scale=0.45]{orcid}}\,$^{1}$, \\
Luis Quiroga\href{https://orcid.org/0000-0003-2235-3344}{\includegraphics[scale=0.45]{orcid}}\,$^{1}$, and Fernando J. G\'omez-Ruiz\href{https://orcid.org/0000-0002-1855-0671}{\includegraphics[scale=0.45]{orcid}}\,$^{2,*}$}
\date{}
\maketitle
\vspace{-1cm}
\begin{center}
$^{1}${\it Departamento de F{\'i}sica, Universidad de los Andes, A.A. 4976, Bogot\'a D. C., Colombia}\\
$^{2}${\it Instituto de F\'isica Fundamental IFF-CSIC, Calle Serrano 113b, Madrid 28006, Spain}\\
$^{*}${\it Corresponding Author:} \href{mailto:fernandojavier.gomez@iff.csic.es}{fernandojavier.gomez@iff.csic.es}
\end{center}
\begin{abstract}
The Kibble-Zurek mechanism (KZM) captures the essential physics of nonequilibrium quantum phase transitions with symmetry breaking. KZM predicts a universal scaling power law for the defect density which is fully determined by the system's critical exponents at equilibrium and the quenching rate. We experimentally tested the KZM for the simplest quantum case, a single qubit under the Landau-Zener evolution, on an open access IBM quantum computer (IBM-Q). We find that for this simple one-qubit model, experimental data validates the central KZM assumption of the adiabatic-impulse approximation for a well isolated qubit. Furthermore, we report on extensive IBM-Q experiments on individual qubits embedded in different circuit environments and topologies, separately elucidating the role of crosstalk between qubits and the increasing decoherence effects associated with the quantum circuit depth on the KZM predictions. Our results strongly suggest that increasing circuit depth acts as a decoherence source, producing a rapid deviation of experimental data from theoretical unitary predictions.\\
\\
{\small\bf Keywords: IBM Quantum Computing, Kibble-Zurek Mechanism, Landau-Zener Model, Adiabatic-Impulse approximation, Quantum Technologies}
\end{abstract}

Characterizing the non-equilibrium dynamics in noisy intermediate-scale quantum (NISQ) devices plays an important role in developing both hardware and architecture designs in the search for scalable quantum computers. NISQ devices have recently attracted tremendous interest, resulting in rapid progress in fundamental studies of novel hardware and architecture together with promising potential for quantum computing~\cite{Preskill2018, Kishor_RMP22}. For example, advancements in NISQ devices demonstrate a ``quantum advantage" in solving sampling problems~\cite{Arute_Nat19, Han_Sci20, Mooney_AQT21}. To further improve quantum advantage, it is desirable that devices show important features such as high-fidelity gates, qubits with long coherence times, control of state preparation and measurement.~\cite{Flammia_PRL11, daSilva_PRL11, Proctor_PRL19}. Open-access/online NISQ devices have recently become readily available, such as those provided publicly by the IBM Quantum Experience platform~\cite{IBMQ2}, showing a significant improvement in the last few years. Despite suffering from noise and scalability limitations, this platform offers a unique possibility to experiment with actual few qubit quantum devices in order to carry out a rigorous study of dynamical quantum properties in different settings along the real time-dynamics of quantum hardware.\\

A key feature of merit in the current NISQ regime is the ability to simulate non-equilibrium quantum dynamics. The Kibble-Zurek mechanism (KZM)~\cite{Kibble76a, Kibble76b, Zurek96a, Zurek96c} is a prominent paradigm to unravel signatures of universal dynamics in the scenario of a finite-rate spontaneous symmetry breaking. The KZM predicts the production of topological defects (kinks, vortices, strings) or in general, non-equilibrium excitations (in both short- and large-ranged interacting systems) in the course of either quantum~\cite{ZDZ05, Dziarmaga05,AcevedoPRL} or classical~\cite{Kibble76b, Zurek96a} phase transitions. The key result of KZM is concerned with the fact that the mean value of density of topological defects scales as a power law of the quench rate. Furthermore, new evidence of scaling in the high-order cumulants has also been recently shown~\cite{Fernando20,delcampo18}. These theoretical predictions have been observed in various experimental platforms such as Bose Gas~\cite{Goo_PRL20}, trapped ions~\cite{Cui_2020}, quantum annealer~\cite{Bando_PRR20, King_Arx22}, Bose-Einstein Condensate~\cite{Anquez_PRL16,DamskiPRL07}, and Rydberg atoms~\cite{Keesling_Nat19}.\\
\\
Damski et al.~\cite{Damski_PRL05,Damski_PRA06,DamskiPRA07} established a close relationship between second order quantum phase transitions and avoided level crossing evolutions, thus establishing the Landau-Zener (LZ) model itself as the simplest paradigmatic scenario for probing KZM~\cite{Landau_32a,Landau_32b,Zener_PRS32,Stukelberg32,Majorana_PRS32}. The density of topological defects can be expressed as a transition probability for a two-level system. Therefore, this relationship can be tested in generic single qubit platforms. This relationship has been probed by using optical interferometry~\cite{Xiao_PRL14}, superconducting qubits~\cite{Gong_SP16, Wang_PRA14} and trapped ion systems~\cite{Cui_2016}.\\
\\
IBM-Q currently grants access up to 5-qubit quantum machines based on superconducting transmon qubits which are controllable using {\it Qiskit}, an open-source software development kit~\cite{Qiskit1, Qiskit2}. These machines have been successfully utilized in simulating spin models~\cite{Cervera_Q18, Rodriguez_Vega_2022}, topological fermionic models~\cite{Koh_2022}, quantum entanglement~\cite{Pozzobom_2019,Wang_2018,Choo_2018,Cruz_2019,Mooney_2019}, far-from-equilibrium dynamics~\cite{Zhukov_2018}, non-equilibrium quantum thermodynamics~\cite{Solfanelli_PRXQ21,Gherardini_PRA21}, open-quantum systems~\cite{Guillermo_NPJ20}, among others. One of the future advantages of IBM-Q is the possibility to do simulation of quantum systems beyond the maximum limits of classical computer over a wide range of parameters. In this work, we test the KZM adiabatic-impulse assumption on the simplest, but important case of a single qubit (LZ model), through experiments on the Qiskit~\cite{Qiskit1} simulator and real quantum hardware, establishing the limits required to obtain accurate results in each case. We successfully reproduced the LZ dynamics under a discrete time evolution in current IBM quantum devices which can provide information about dynamics state evolution  given that error mitigation procedures were implemented. Additionally, noticeable effects of decoherence are observed and explained by a simple phenomenological model of relaxation and dephasing for open quantum systems. Furthermore, analysis and estimation of the experimental asymptotic probability allows us to verify the universal KZM in a timescale appropriate for an almost closed system under an adiabatic quench regime. In summary, the key achievement of this work has been the validation of a central premise of KZM through a protocol to characterize and obtain an effective time-dependent dynamics on IBM realistic quantum computers. For reaching such goal we performed LZ evolution under different annealing times, maintaining a fixed number of total gates, a basic benchmark procedure on quantum critical phenomena in near term quantum computers.\\
\\
This paper is organized as follows. A brief review on KZM, the LZ model and its close connection with KZM are presented in Sect.~\ref{section2}. In Sect.~\ref{section3} we present the experimental platform. The contrast between theoretical predictions and experimental results is collected in Sect.~\ref{section4}. Finally, we summarize the main conclusions in Sect.~\ref{Conclu}.

\section{Theoretical background}\label{section2}
\subsection{Brief review of the Kibble-Zurek Mechanism}
\begin{figure}[t]
\includegraphics[width=1\linewidth]{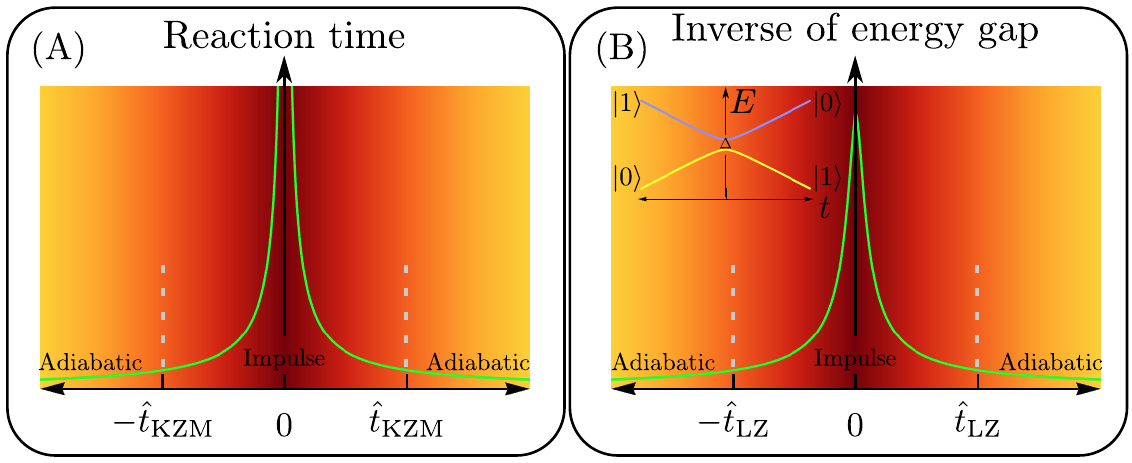}
\caption[justification=justified]{\label{fig1}{\bf Connection between KZM and avoided level crossing in a LZ transition.} (A)
In a continuous second order phase transition, the reaction time diverges near the critical point. The KZM approximation takes into account the total dynamics divided in three stages (adiabatic, impulse, and adiabatic) represented by the graduated  yellow-dark red-yellow colors and separated by the freeze out-time $\hat{t}_{{\rm KZM}}$. (B) The inverse of the energy gap in LZ exhibits a similar behavior of the reaction time. However, it is not divergent at the crossing point. Similarly, we divided the LZ dynamics in the same three KZM regimes and separated by the Landau-Zener jump time $\hat{t}_{{\rm LZ}}$. Inset: Avoided level crossing LZ.}
\end{figure}

The KZM describes the dynamics of a system across a continuous symmetry breaking second-order phase transition induced by the change of a control parameter $\lambda$. When the system is driven through the critical point $\lambda_c$, both the correlation length $\xi$ and reaction time $\tau$ diverge as
\begin{equation}\label{div_1}
\xi=\xi_0\left|\epsilon\right|^{-\nu}, \quad\tau = \tau_0 \left|\epsilon\right|^{-z\nu}.
\end{equation}
where, $\epsilon=\pap{\lambda - \lambda_c}/\lambda$ marks the separation from the critical point. The spatial and dynamic {\it equilibrium} critical exponents are given by $\nu$ and $z$, respectively, while the mesoscopic behavior of the system is contained in the dimensional constants $\xi_0$ and $\tau_0$. If the quench varies linearly in time, $\epsilon\pap{t}=t/t_a$, where $t_a$ denotes a quench or annealing time scale, the system reaches the critical point at $t=0$.  Therefore, the equilibrium effective reaction time diverges as Eq.~\eqref{div_1}. This phenomenon is known as critical slowing down and can be used to describe the time evolution across a phase transition as a sequence of three stages. Initially, the system is prepared in the high symmetry phase from which it evolves within an adiabatic evolution stage. Secondly, the evolution enters an impulse stage in the neighborhood of the phase transition where the system is effectively frozen. Finally, when the system is far away from the critical point, the dynamics are adiabatic again. These three regimes are schematically represented in Fig.~\ref{fig1}(A). The three regions are separated by two points marked as $-\hat{t}_{{\rm KZM}}$ and $\hat{t}_{{\rm KZM}}$, in such a way that the freeze-out occurs at the instant $\hat{t}_{{\rm KZM}}\sim\pap{\tau_0 t_a^{z\nu}}^{1/1+z\nu}$. The main point of the KZM argument is that the size average or correlation length, $\hat{\xi}$, of domains in the broken symmetry phase is set by the equilibrium correlation length evaluated at the freeze-out time. Therefore, the density of excitations resulting from quench evolution scales as $\rho\sim\hat{\xi}^{-D}$ and goes as
   \begin{equation}
   \rho_{\text{KZM}}\sim\frac{1}{\xi_0}\pap{\frac{\tau_{0}}{t_a}}^{\frac{D\nu}{1+z\nu}},
    \end{equation}
where $D$ is the dimensionality of the system. This result was initially derived in the classical domain~\cite{Kibble76b, Zurek96a} and subsequently extended to quantum systems~\cite{ZDZ05, Dziarmaga05}. Additionally, \blue{t}he KZM has also been extended to novel scenarios including long-range interactions~\cite{PueblaPRA1,PueblaPRL,AcevedoPRL}, inhomogeneous systems~\cite{FernandoPRL19,DM10,ColluraPRL10}, and nonlinear quenches~\cite{SenPRL08,BarankovPRL08}.
\subsection{Landau-Zener model}\label{sec_LZ}
 Consider a two-level system, with gap $\Delta$, described by the time-dependent Hamiltonian $\pap{\hbar = 1}$
\begin{equation}\label{LZ_Hamil}
\hat{H}\pap{t}=-\frac{\varepsilon\pap{t}}{2}\hat{\sigma}_{z}-\frac{\Delta}{2}\hat{\sigma}_x.
\end{equation}
With $\hat{\sigma}_{n}$ the Pauli matrix along the $n\in\pac{x,y,z}$ direction. We define the {\it diabatic} states as the Hamiltonian eigenvectors when $\Delta=0$ and consequently eigenvectors for the Pauli operator $\hat{\sigma}_z$: $\hat{\sigma}_z\ket{0}=+1\ket{0}$ and $\hat{\sigma}_z\ket{1}=-1\ket{1}$. The respective (diabatic) energy levels are $E_{0,1}=\mp \varepsilon\pap{t}/2$. Now, the adiabatic instantaneous eigenvalues $E_{\pm}\pap{t}$ and eigenstates $\ket{E_{\pm}\pap{t}}$ are solutions of $\hat{H}\pap{t}\ket{E_{\pm}\pap{t}}=E_{\pm}\pap{t}\ket{E_{\pm}\pap{t}}$. The instantaneous gap energy is given by $\Delta E = E_{+}-E_{-}=\sqrt{\varepsilon^2\pap{t}+\Delta^2}$ (for more details see Ref.~\cite{NoriRev}). In the main panel of Fig.~\ref{fig1}(B), we depicted the inverse of the energy gap as a function of time while the instantaneous adiabatic eigenvalues are shown as an inset in the Fig~\ref{fig1}(B). The eigenstates are written as a linear combination of the diabatic states as $\ket{\psi\pap{t}}=\alpha\pap{t}\ket{0}+\beta\pap{t}\ket{1}$. By solving the corresponding eigenequation in terms of parabolic cylinder functions $\mathbf{D}_{p}\pap{z}$, and using the substitution $z=t\exp\pas{i\pi/4}/\sqrt{t_a}$, we obtain the transition amplitudes
\begin{equation}\label{proba1}
\begin{split}
\alpha\pap{z}&=\frac{e^{-i\frac{3\pi}{4}}}{\sqrt{\delta}}\left[\delta\chi_1\mathbf{D}_{-1-i\delta}(z) + \chi_2\mathbf{D}_{i\delta}(iz)\right],\\
\beta\pap{z}&= \chi_1 \mathbf{D}_{-i\delta}\pap{z} + \chi_2 \mathbf{D}_{-1+i\delta}\pap{iz}.
\end{split}
\end{equation}
where $\delta = \Delta^2 t_a/4$ is the adiabaticity parameter. Moreover, $\chi_1$ and $\chi_2$ are found from the initial condition at $z=z_i$ (see the section: Supplemental Data for details of the calculations and derivations):
\begin{equation}\label{init_conf1}
\begin{split}
\chi_1 &= \frac{e^{i\frac{3\pi}{4}}\sqrt{\delta} \mathbf{D}_{-1+i\delta}\pap{i z_i}\alpha\pap{z_i}-\mathbf{D}_{i\delta}\pap{i z_i}\beta\pap{z_i}}{\delta \mathbf{D}_{-1-i\delta}\pap{z_i}\mathbf{D}_{-1+i\delta}\pap{i z_i} -\mathbf{D}_{-i\delta}\pap{z_i} \mathbf{D}_{i\delta}\pap{i z_i}},\\
\chi_2 &= \frac{-e^{i\frac{3\pi}{4}}\sqrt{\delta}\mathbf{D}_{-i\delta}(z_i) \alpha\pap{z_i} + \delta\mathbf{D}_{-1-i\delta}(z_i)\beta(z_i)}{\delta \mathbf{D}_{-1-i\delta}\pap{z_i}\mathbf{D}_{-1+i\delta}\pap{i z_i} -\mathbf{D}_{-i\delta}\pap{z_i} \mathbf{D}_{i\delta}\pap{i z_i}}.
\end{split}
\end{equation}
Notice that, Eqs.~\eqref{proba1} and~\eqref{init_conf1} are valid for any arbitrary initial condition and final time $t$. For the experimental implementation discussed below, we are interested in studying the system's evolution from an initial state starting in the anticrossing point at $t=0$. In the section: Suplemental Data, the formal solutions for this particular initial condition are summarized.
\subsection{Connection between the KZM and LZ evolution}
Here we demonstrate how we can implement a controllable evolution using an IBM-Q quantum simulation, in close analogy to the topological defect formation in KZM. Following the seminal arguments exposed in Ref.~\cite{Damski_PRL05,Damski_PRA06}, topological defects can be built into the LZ model by being associated to the diabatic states. Consider one of the states, such as $\ket{0}$, to be a topologically defected phase and $\ket{1}$ a defect-free phase. For example, in the case of vortices, state $\ket{0}$ may be an eigenstate of the angular momentum operator $\hat{L}_z\ket{0} = n\ket{0}$, while $\hat{L}_z\ket{1} = 0$. In this scenario, Damski introduces the normalized density of topological defects as the average angular momentum
\begin{equation}
    \rho_{{\rm KZM}} = \frac{1}{n}\bra{\psi}\hat{L}_z\ket{\psi} = \pabs{\langle\psi|0\rangle}^2.
\end{equation}
Then, a system evolving in time under the LZ model can be used to study transitions between the phases through the probabilities of the diabatic states. The similarity between the reaction time of a second order phase transition and the inverse of energy gap in the LZ Hamiltonian is shown in Fig~\ref{fig1}. In analogy with the KZM, this suggests that the adiabatic-impulse-adiabatic approximation (AI) may be used to estimate the asymptotic probability when the system traverses the avoided level crossing, thus elucidating the link between the KZM and LZ evolution.\\
\\
We divided the dynamics through the anti-crossing into three stages like the AI scenario for KZM. Without loss of generality, we assume that the system starts at $t_i\to -\infty$ from the ground state $\ket{E_{-}}$, and then it evolves to $t_f \to \infty$. We define a natural time scale given by the inverse of the energy gap
\begin{equation}
\frac{1}{E_{+}\pap{\hat{t}_{{\rm LZ}}}-E_{-}\pap{\hat{t}_{{\rm LZ}}}}=\eta \hat{t}_{{\rm LZ}}\blue{,}
\end{equation}
where $E_{\pm}\pap{t}$ are the adiabatic energy eigenvalues at time $t=\hat{t}_{{\rm LZ}}$ and $\eta$ is a constant. Using Eq.~\eqref{LZ_Hamil}, we obtain
\begin{equation}
\frac{\hat{t}_{{\rm LZ}}}{t_a}=\frac{\Delta}{\sqrt{2}}\sqrt{\sqrt{1 +\frac{4}{ \pap{\Delta^2 \eta  t_{a}}^2}}-1}.
\end{equation}
The AI assumes that the evolution wave function $\ket{\psi\pap{t}}$ of the system satisfies:
\begin{itemize}
\item Adiabatic dynamics: from $t_i=-\infty$ to $t=-\hat{t}_{{\rm LZ}}$
\[\ket{\psi\pap{t}}\approx e^{i\Phi_1} \ket{E_{-}\pap{t}}.\]
\item Impulse dynamics: from  $t=-\hat{t}_{{\rm LZ}}$ to $t=\hat{t}_{{\rm LZ}}$
\[\ket{\psi\pap{t}}\approx e^{i\Phi_2} \ket{E_{-}\pap{-\hat{t}_{{\rm LZ}}}}.\]
\item Adiabatic dynamics: from $t=\hat{t}_{{\rm LZ}}$ to $t_f = \infty$
\[\pabs{\langle\psi\pap{t}\ket{E_{-}\pap{t}}}^2\approx A.\]
\end{itemize}
Where $\Phi_1$, $\Phi_2$ are global phases, and $A$ is a constant. Following the AI, Damski in Refs.~\cite{Damski_PRL05,Damski_PRA06} reported the probability of finding the LZ system in the excited state at $t_f \gg t_{{\rm LZ}}$, a calculation we briefly summarize for the sake of completeness in view of our main experimental in
terest.\\
\\
From now on, we focus on the LZ dynamics for the evolution starting in the ground state at the anticrossing point. The initial state at $t=0$ is then expressed as $\ket{E_{-}\pap{0}}=\pap{\ket{1}-\ket{0}}/\sqrt{2}$, and consequently the transition probability $P_{{\rm AI}} =\pabs{\langle E_{+}\pap{\hat{t}_{{\rm LZ}}}\ket{E_{-}\pap{0}}}^2$ is given by~\cite{Damski_PRL05,Damski_PRA06}
\begin{equation}\label{PAI_1}
P_{{\rm AI}} =\frac{1}{2}\pap{1-\frac{1}{\sqrt{1+\hat{\varepsilon}^2}}}= \frac{1}{2}-\frac{1}{2}\sqrt{1-\frac{2}{\pap{\eta t_a}^2+\eta t_a\sqrt{\pap{\eta t_a}^2 +4}+2}}.
\end{equation}
Where we have fixed the two-level system gap to $\Delta=1$. Additionally, $\hat{\varepsilon}=\varepsilon\pap{\hat{t}_{{\rm LZ}}}$ is the linear bias at time $t=\hat{t}_{{\rm LZ}}$. Expanding Eq.~\eqref{PAI_1} into a series of $\sqrt{t_a}$, we obtained~\cite{Damski_PRL05,Damski_PRA06}
\begin{equation}\label{AI_Series1}
P_{{\rm AI}}=\frac{1}{2}-\frac{\sqrt{\eta}}{2}t_a^{1/2}+\frac{\eta \sqrt{\eta}}{8}t_a^{3/2}+\mathcal{O}\pap{t_a^{5/2}}.
 \end{equation}
which will be relevant for testing the predictions of the universal AI for KZM below.

\begin{figure}[t]
\centering
\includegraphics[width=1\linewidth]{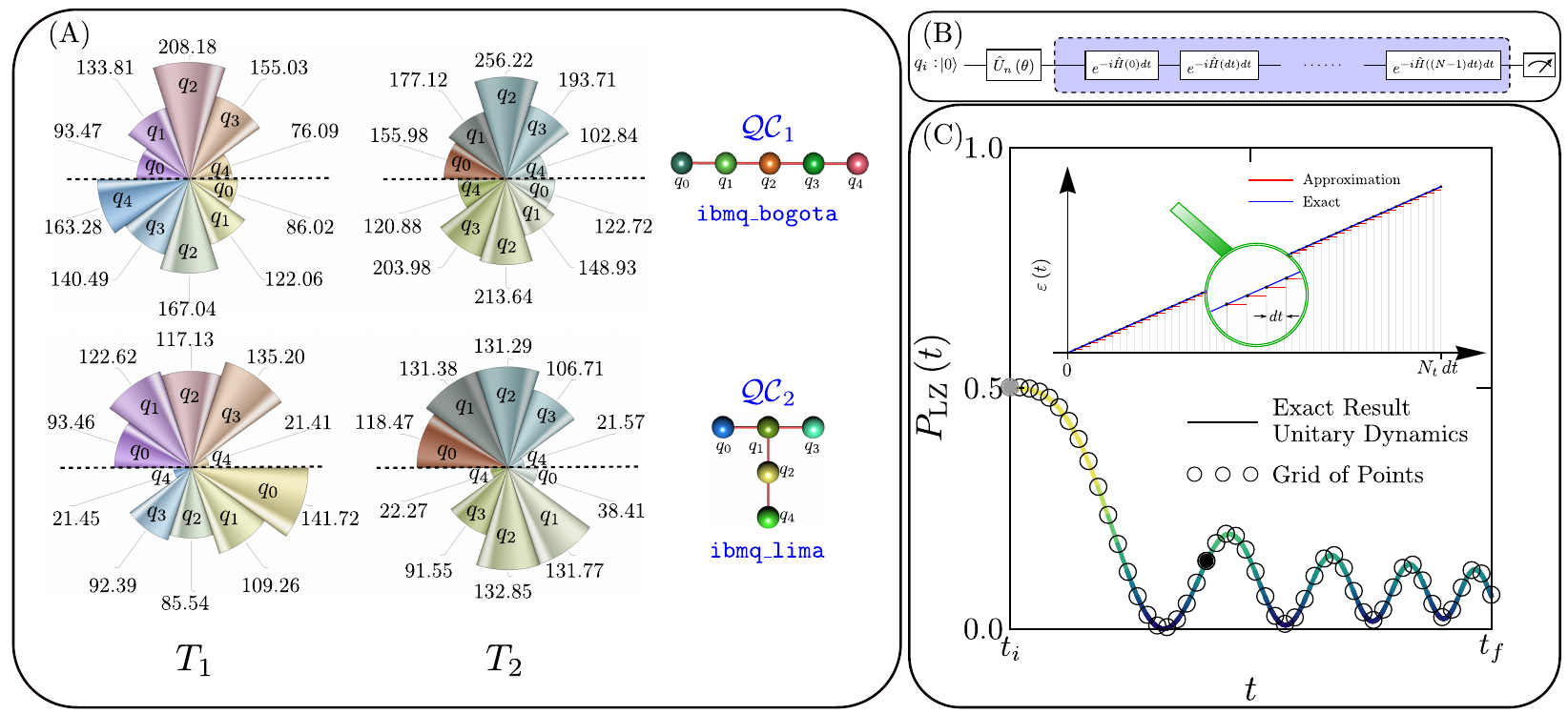}
\caption[justification=justified]{\label{fig2A}{\bf Decoherence times in different IBM-Q and IBM-Q circuit simulation of the Landau-Zener process.} (A) In the pie-like chart, we contrast the thermal relaxation time $(T_1)$ and dephasing time $(T_2)$, in $\pap{\mu s}$, for two different topology circuits, simply called $\mathcal{QC}_1$ and $\mathcal{QC}_2$ (see text for details). Due to in situ IBM machine calibration routines the times $T_1$ and $T_2$ may change. Every pie-like chart is divided into two sectors by a dashed line, where the upper and lower sectors corresponding to decoherence times at two different dates. (B) Quantum circuit for the LZ simulation starting at the state $\ket{\psi\pap{t_i}}=\hat{U}_{n}\pap{\theta}\ket{0}$, where $\hat{U}_{n}\pap{\theta}$ is an unitary rotation along the axis $n$. (C) Schematic representation of the LZ transition probability: the solid line corresponds to the exact result given by Eq.~\eqref{Prob_LZ}, with  $\Delta=1$, $t_a = 2$, $t_i = 0$ and $t_f=10$, while the symbols illustrate expected results for a grid of points with separation $dt =t_f -t_i/N_t$, being $N_t$ the total circuit depth. The filled dots correspond to: the shortest circuit with depth $1$ (gray dot) and an intermediate circuit depth $N$ (black dot). The inset shows the discrete approximation of the time-dependent component $\varepsilon(t)$ of the LZ Hamiltonian.}
\end{figure}

\section{Experimental IBM-Q platform}\label{section3}
We implemented our experimental studies in two topologies or processors types. Figure~\ref{fig2A}(A) shows the device layout for the IBMQ 5-qubit \texttt{ibmq\_bogota} (Falcon r5.11L topology $\mathcal{QC}_1$) and \texttt{ibmq\_lima} (Falcon r4T topology $\mathcal{QC}_2$). The topology of the device determines the possible placement of two-qubit gates. The qubits are furthermore prone to decoherence, thereby requiring several runs of the experiment to make up for statistical errors. We measure
the LZ, and concomitant KZM relation, for each one of the IBM-Q transmons in $\mathcal{QC}_1$ and $\mathcal{QC}_2$. Each transmon
plays the role of a qubit, evolving with its own dynamics, experimentally showing the effects of decoherence
on the hardware. Generally, the physical transmon type qubits of the same machine offer a variety of properties that describe the quality of the qubit, such as thermal relaxation time $(T_1)$, dephasing time $(T_2)$, anharmonicity, and error properties detailed in the section: Suplemental Data, allowing us to compare the simulation's performance with different physical parameters.  In Fig.~\ref{fig2A}(A), the times $T_1$ and $T_2$ are depicted for each considered circuit topology at two different dates, illustrating in a graphical way how these times change every time that IBM performed a calibration of every device.
\section{Results}\label{section4}
\subsection{Simulation of the Landau-Zener evolution on IBM-Q}
{\it Unitary dynamics.--} We are interested in the experimental determination, and respective simulation, on a digital open-access IBM-Q of a single qubit evolution under a linearly time-dependent Hamiltonian (LZ problem). At time $t_i$, a qubit in the processor is initialized in the state $\ket{\psi\pap{t_i}}=\hat{U}_{\hat{n}}\pap{\theta}\ket{0}$, where $\hat{U}_{n}\pap{\theta}=\cos\pas{\theta/2}\hat{I}-i\hat{\sigma}_{n}\sin\pas{\theta/2}$ is a unitary rotation along the axis $n\in\pac{x,y,z}$ with $\hat{\sigma}_{n}$ the usual Pauli matrix along the $n$-direction. The whole evolution from $t_i$ to $t_f$ is performed by sampling the Hamiltonian at regular intervals $dt = (t_f-t_i)/N_t$ where $N_t$ denotes the number of time steps or the total circuit depth (see blue region in Fig.\ref{fig2A}(B)). The equivalent circuit for the experimental IBM-Q realization, and its simulation, is shown in Figure~\ref{fig2A}(B). Assuming an evolution governed by a time-independent Hamiltonian and for small enough intervals of duration $dt$, the time evolution operator at time $t=N\>dt$, with $1\leq N \leq N_t$, can be approximated by
\begin{equation}\label{Approximate_TimeEvolutionOperator}
\hat{U}\pap{t,t_i}\approx\prod_{k=0}^{N-1}e^{-i\hat{H}_kdt}\blue{,}
\end{equation}
where $\hat{H}_k=\hat{H}(t_i+kdt)$.\\
\\
Since we are interested in the evolution from an initial condition where the LZ system is prepared in an equal weight superposition at the anticrossing point, we start by applying the unitary rotation $\hat{U}_{y}\pap{-\pi/2}$. The approximate time evolution operator is constructed with general unitary gates. A general unitary single qubit gate describes rotations on the Bloch sphere and is defined by three Euler angles
\begin{equation}
    \hat{U}(\theta,\phi,\lambda) = \begin{pmatrix}\cos\pap{\frac{\theta}{2}} & -e^{i\lambda}\sin\pap{\frac{\theta}{2}}\\[1ex]
    e^{i\phi}\sin\pap{\frac{\theta}{2}} & e^{i\pap{\phi+\lambda}}\cos\pap{\frac{\theta}{2}}\end{pmatrix}.
\end{equation}
IBM-Q devices are equipped with the finite and complete set $\{CX, I, U_z, \sqrt{X}, X\}$ of basis gates on which any quantum circuit must be decomposed into. The general unitary gate can then be expressed using the previous set as
$U(\theta,\phi,\lambda) = e^{i\gamma}\hat{U}_z\pap{\phi+\pi}\sqrt{X}\hat{U}_z\pap{\theta+\pi}\sqrt{X}\hat{U}_z\pap{\lambda}$, where $\gamma = \pap{\lambda + \phi + \pi}/2$ is a global phase factor. Using this decomposition, small time progressions as defined in Eq.~\eqref{Approximate_TimeEvolutionOperator} are simulated and finally the state $\ket{\psi\pap{t}}$ is measured.\\
\\
As already stated, the Landau-Zener dynamics can be exactly solved (see Supplementary Material), thus allowing a direct benchmark test of the experimental results on a realistic quantum device against exact results. For a LZ evolution starting at the anticrossing ground state, we obtain the the LZ transition probability $P_{{\rm LZ}}(t)$ at time $t$ given as
\begin{equation}\label{Prob_LZ}
P_{{\rm LZ}}\pap{t}=\lvert \chi_1 \mathbf{D}_{-i\delta}\pap{z} + \chi_2 \mathbf{D}_{-1+i\delta}\pap{iz}\rvert^2\blue{,}
\end{equation}
with the amplitudes $\chi_1$ and $\chi_2$, see Eq.~\eqref{init_conf1}, given by:
\begin{align}
\chi_1 &= -\frac{2^{k}\exp\pas{i\pi k}}{4\sqrt{ik}}\pas{\frac{\sqrt{2ik}\Gamma\pap{k}+\pap{1+i}\Gamma\pap{\frac{1}{2}+k}}{\Gamma\pap{2k}}},\\
\chi_2&= \frac{\exp\pas{i\pi k}}{2^{k+1}}\pas{\frac{2ik\Gamma\pap{\frac{1}{2}-k}+\pap{1-i}\sqrt{2ik}\Gamma\pap{1-k}}{\Gamma\pap{1-2k}}}.
\end{align}
where $z$ and $\delta$ are given in Sec.~\ref{sec_LZ}.\\
\\
Our first aim is to benchmark our LZ experimental results with the above exact theoretical prediction. This is schematically illustrated in Fig.~\ref{fig2A}(C) where we display the exact result, see Eq.~\eqref{Prob_LZ}, and a hypothetical grid of points representing expected target data with a separation $dt =(t_f -t_i)/N_t$, being $N_t$ the total circuit depth. For every experimental data, 5000 shots have been realized on each quantum circuit, $\mathcal{QC}_1$ and $\mathcal{QC}_2$.\\

\begin{figure}[t!]
\includegraphics[width=1\linewidth]{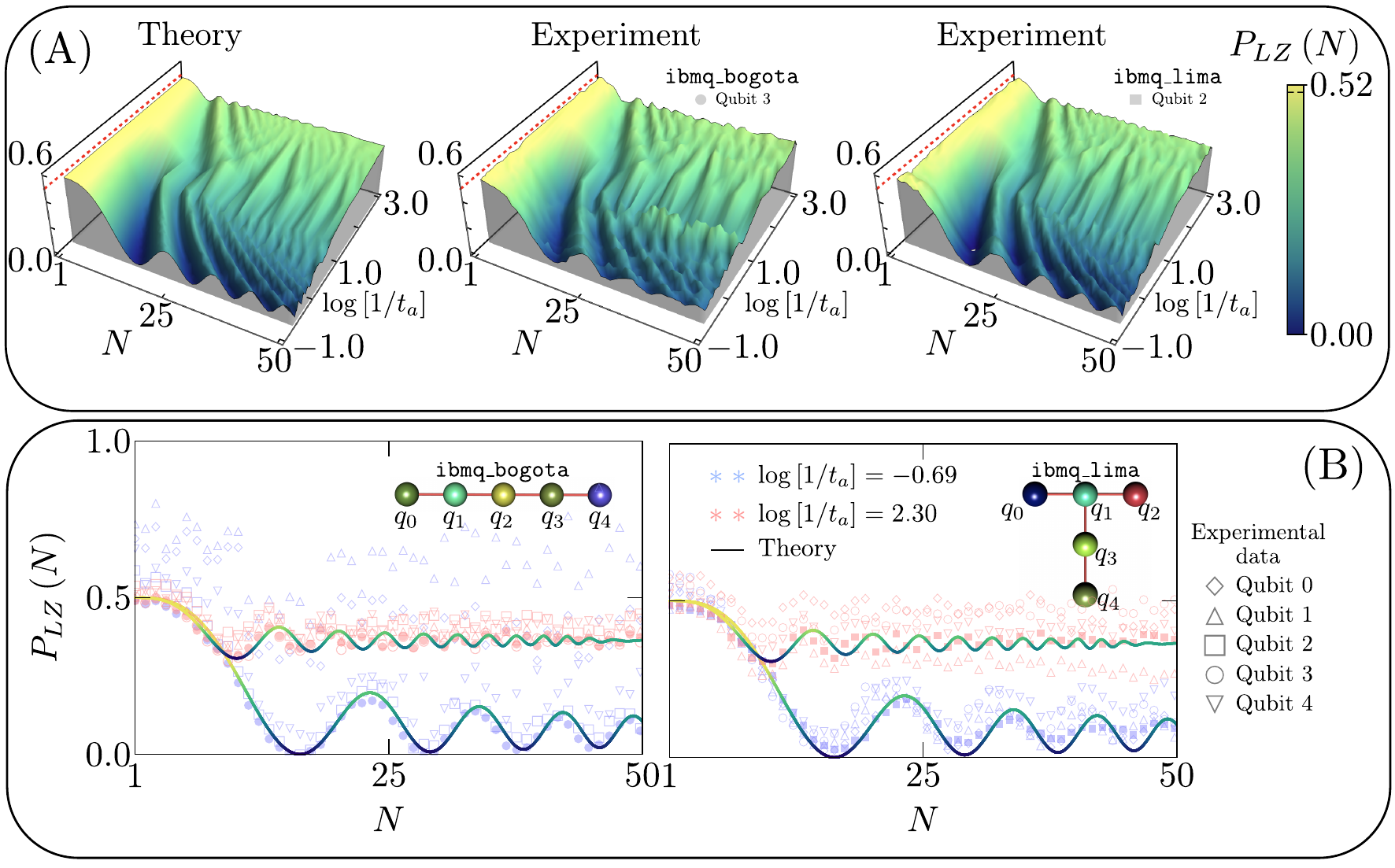}
\caption[justification=justified]{\label{fig3}{\bf Measurement of LZ probabilities on IBM-Q.} In panel (A), we establish a contrast between the exact and experimental results for the LZ transition probability as a function the number of layers or circuit depth $N$, and the annealing time $t_a$. In this panel, all figures share the same color vertical scale. The initial condition, $P_{{\rm LZ}}\pap{t=0}=0.5$, is represented by a red dashed line. Note that in some region of parameters a probability larger than $0.5$ for the experimental results is obtained. In panel (B), the behavior of the LZ probability for every qubit available in each processor is shown, identifying in this way the most isolated (larger decoherence time) qubit in each case. We fixed the maximum number of layers in the circuit as $N_t=50$.}
\end{figure}

\begin{figure}[t!]
\includegraphics[width=1\linewidth]{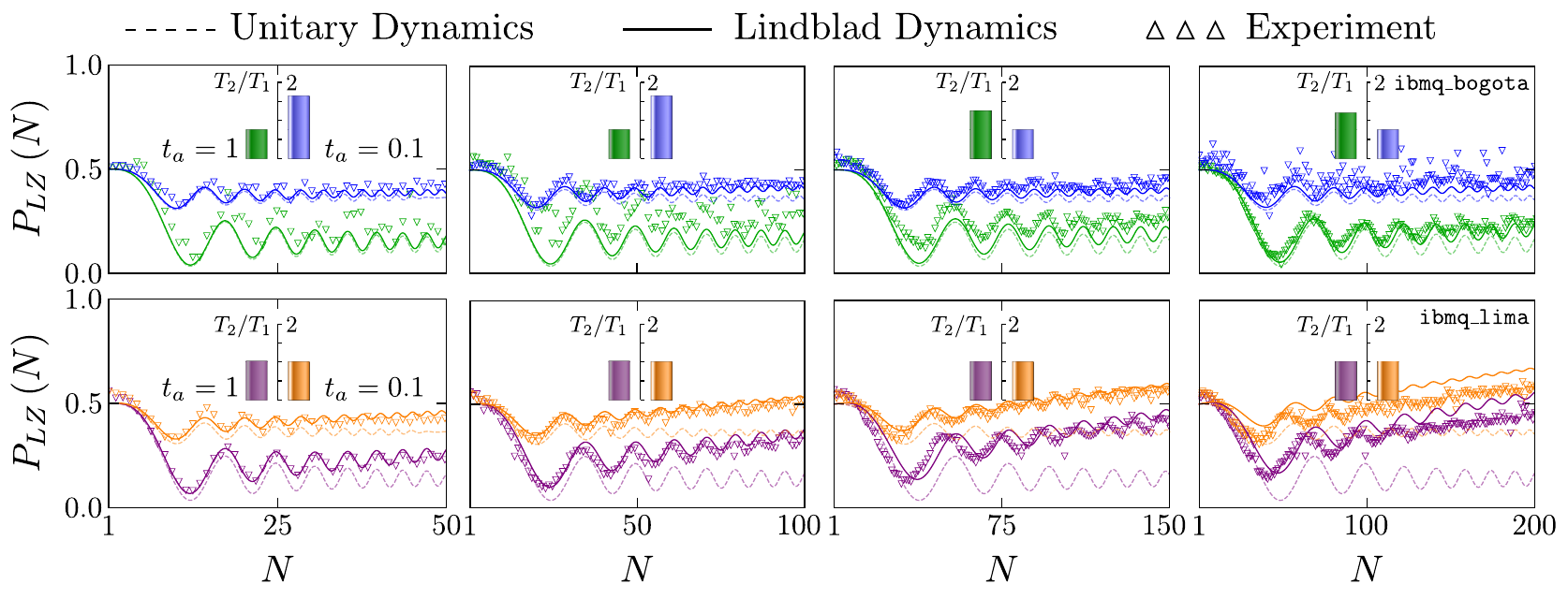}
\caption[justification=justified]{\label{fig5}{\bf Contrast between close and open quantum dynamics for LZ on IBM-Q.} The Landau-Zener transition probability is shown as a function of the number of layers $N$ in the circuit implemention for qubit 4, the noisiest qubit for each $\mathcal{QC}_1$ and $\mathcal{QC}_2$ quantum circuit. We contrast the theoretical prediction for a close system (unitary dynamics) given by Eq.~\eqref{Prob_LZ} (dashed line), the decoherent dynamics given by the numerical solution of the Lindblad equation (solid line) and experimental results (symbols). The experimental results clearly depart from the unitary evolution prediction as the number of layers $N_t$ increases in the circuit (see Fig.~\ref{fig2A}(B) blue region). Additionally, in every panel, we present as inset the ratio between the dephasing time $(T_2)$ and the thermal relaxation time $(T_1)$.}
\end{figure}

The unit of energy is set by choosing $\Delta =1$ in the LZ Hamiltonian (see Eq.~\eqref{LZ_Hamil}). Therefore, in the following, we express energy parameters and time as dimensionless quantities ($\hbar=1$). Using the quantum circuits $\mathcal{QC}_1$ and $\mathcal{QC}_2$, we implemented the corresponding gates in all qubits available on parallel and we did a sweep of parameters in annealing time $t_a$ from $0.05$ to $2.0$. Additionally, for both theoretical and experimental results, the final evolution time $t_f$ was chosen according with: $t_f =4$ for annealing times in the interval $0.05\leq t_a \leq 0.17$ and $t_f =10$ for $0.17< t_a \leq 2$. These particular choices have been supported by the fact that as we are mainly interested in the asymptotic LZ probability transition, a good asymptotic collapse is reached for these parameter regimes. We also represent the experimental results $P_{{\rm LZ}}\pap{N}$ as a function of the number of layers in the circuit instead of time. We emphasize that an $N$-deep circuit corresponds to a physical qubit interaction time $t_{{\rm Int}} = 2 t_{SX} N$,  where $t_{SX}$ is the gate length property for $\sqrt{X}$ and it is fixed by IBM-Q as $t_{SX}= 35.555\text{ns}$. In Fig.~\ref{fig3}, we present a contrast of the LZ transition probability for both the theoretical and experimental results. In the panel Fig.~\ref{fig3}(A), we choose the most robust qubit that better reproduced the theoretical $P_{{\rm LZ}}$. Specifically, we found that the qubit $3$ and $2$ for \texttt{ibmq\_bogota} and \texttt{ibmq\_lima}, respectively, have the best performance. In order to better appreciate the experimental agreement and differences for every single-qubit over $\mathcal{QC}_1$ and $\mathcal{QC}_2$, we show the LZ transition probability as a function of the number of applied gates in Fig.~\ref{fig3}(B).\\
\\
In the next subsection, we address the influence of the number of layers in the LZ simulation circuit and the role of decoherence.\\
\\
{\it Open system dynamics.-} The performance of the hardware worsens with an increasing depth of the circuit. The assumption of a closed quantum system rapidly breaks down for qubits with short relaxation ($T_1$) and dephasing ($T_2$) timescales, thus requiring for a theoretical analysis that resorts to a quantum open system approach. The effects of quantum decoherence are noticeable in the measured probability when scaling the number of gates due to the increase in computing times. We model every qubit on IBM-Q as a two-level system coupled to a Markovian bath. The system evolution is described by a continuous map $\rho_t = e^{t\mathcal{L}}\rho_{t_0}$, $t\geq 0$ generated by the Lindbladian $\mathcal{L}\pas{\bullet} =-i\pas{\hat{H},\bullet}+\sum_{n}\pap{\hat{L}_n\bullet \hat{L}_n^\dagger - \frac{1}{2}\pac{\hat{L}_{n}^{\dagger}\hat{L}_n,\bullet}}$~\cite{Breuer_Book}, where, $\hat{H}$ is the Hamiltonian and $\pac{\hat{L}_n}$ are Lindblad operators that describe the system-bath interactions. Dissipative processes in a superconducting qubit such as relaxation, i.e., transitions from the higher energy level $\ket{1}$ to ground state $\ket{0}$, can be described phenomenologically by the operator $\hat{L}_1 = \sqrt{\Gamma}\lvert 0\rangle\langle1\rvert$ and dephasing by rotations around the $z$ axis $\hat{L}_2 = \sqrt{\gamma}\hat{\sigma}_z$. Additional transitions such as thermal excitations from the ground state $\ket{0}$ to $\ket{1}$ may also be considered \cite{introDissipationDecoherence}, although for a superconducting transmon qubit this process is negligible. The rates $\Gamma = 1/T_1$ and $\gamma = 1/T_2-1/2T_1$ are related to the characteristic times of each physical qubit.\\
\\
In Fig.~\ref{fig5}, we establish a contrast between the unitary exact dynamics, numerical Lindblad dynamics (QuTip) and the experimental results obtained for qubit 4, the noisiest qubit in both quantum machines.  QuTiP is an open-source framework for Python that allows for numerical simulations of quantum dynamics of open systems under different solvers~\cite{QuTip1, QuTiP2}. Specifically, we depicted the Landau-Zener probability as a function of the number of layers in the circuit, $N$, for two specific annealing times $t_a = 1$ (colors green/purple) and $t_a = 0.1$ (colors blue/orange). Additionally, we show as an inset the ratio between $T_2/T_1$, the bar scale shows the value of this proportion from 0 to 2. Although, \texttt{ibmq\_lima} quantum computer has the ratio $T_2/T_1$ almost constant, qubit 4 is the most prone to decoherence.

\subsection{Simulation of the Kibble-Zurek mechanism on IBM-Q}
\begin{figure}[t!]
\centering
\includegraphics[width=1\linewidth]{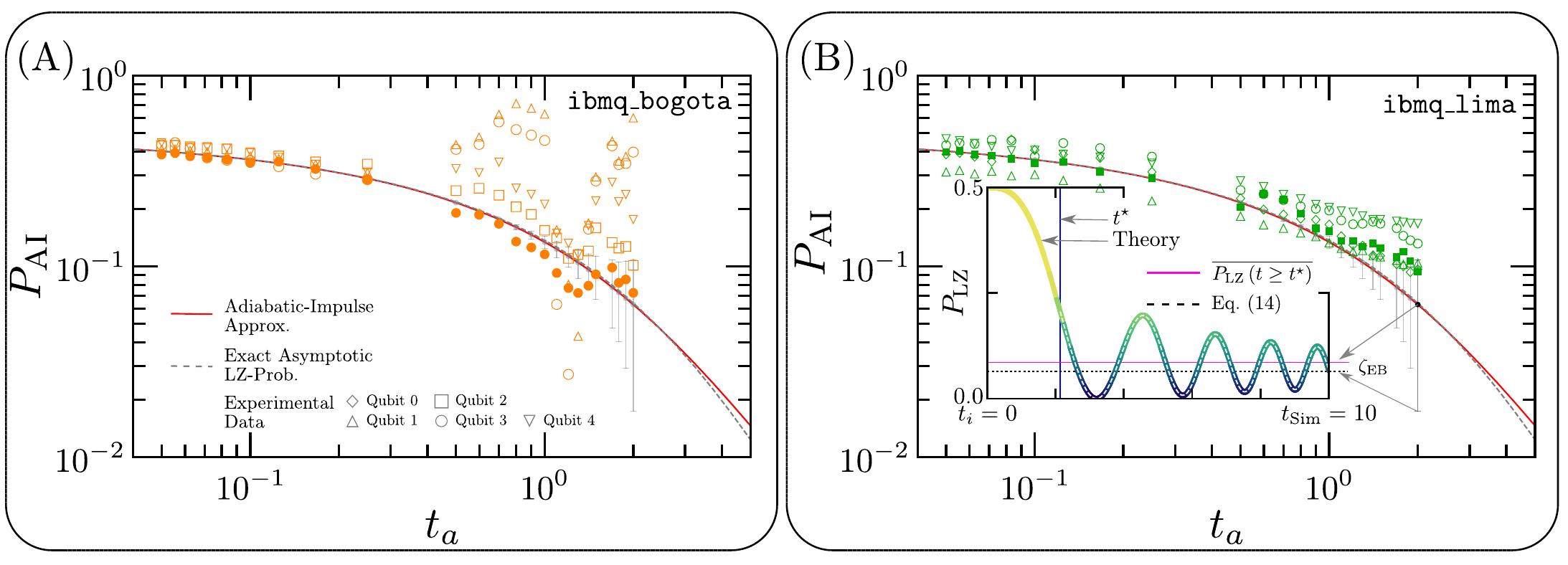}
\caption[justification=justified]{\label{fig4}{\bf Simulation of the Kibble-Zurek mechanism on IBM-Q.} In both upper and lower panels, we contrast the adiabatic-impulse approximation (Eq.~\eqref{PAI_1}), asymptotic Landau-Zener probability (Eq.~\eqref{AsimLZP}), and the experimental data. In panel (a), we show experimental data retrieved from \texttt{ibmq\_bogota}. In panel (b),  we present the experimental results from \texttt{ibmq\_lima}. In the inset, we present the protocol to calculate the asymptotic experimental Landau-Zener probability. The error bars with length $2\zeta_{{\rm EB}}$, calculated from the finite-time effect, are also shown. Solid symbols are consistent with the best qubit behavior as depicted in Fig.~\ref{fig3}.}
\end{figure}

The main purpose of this work is to validate the adiabatic-impulse approximation of the Kibble-Zurek mechanism through the nonequilibrium dynamics of the Landau-Zener model on IBM-Q. Using Eq.\eqref{Prob_LZ} with $\Delta=1$, the asymptotic probability can be exactly calculated as
\begin{equation}\label{AsimLZP}
P_{{\rm LZ}}\pap{t\to\infty}=1-\frac{1}{\delta}\exp\pas{-\frac{3\pi\delta}{2}}|\chi_2|^2.
\end{equation}
Expanding the asymptotic probability into series, we obtain~\cite{Damski_PRL05,Damski_PRA06}
\begin{equation}
P_{{\rm LZ}}\pap{t\to\infty}=\frac{1}{2}-\frac{\sqrt{\pi}}{4}t_a^{1/2}+\frac{\sqrt{\pi}}{32}\pap{\pi-\ln\pap{4}}t_a^{3/2}+\mathcal{O}\pap{t_a^{5/2}}.
\end{equation}
We find the value of $\eta$ by directly comparing the adiabatic-impulse approximation given by Eq.~\eqref{AI_Series1} and the expansion of the LZ asymptotic probability at first-order $(\eta = \pi/4)$. However, non-trivial corrections for high-order terms appear. In both main panels of Figs.~\ref{fig4}(a-b), we depict the agreement of the theoretical prediction for the adiabatic-impulse approximation (Eq.~\eqref{PAI_1}) and asymptotic Landau-Zener probability (Eq.~\eqref{AsimLZP}). We note the role of the corrections for large quench times. For finite-time LZ simulations, estimating the asymptotic transition probability becomes challenging and similar to experimental data. To this end, we introduced the Landau-Zener jump-time $t^{\star}$ as the fist zero in the second derivative of the Landau-Zener probability, thus:
\begin{equation}
\frac{d^2P_{\rm LZ}\pap{t}}{dt^2}\Bigg\vert_{t=t^{\star}}=0.
\end{equation}

In this way, we propose that the estimated finite-time asymptotic Landau-Zener probability can be approximated by the average of all values of $P_{{\rm LZ}}\pap{t}$ with $t\geq t^{\star}$. In the inset of Fig.~\ref{fig4}(B), we display the protocol implemented to calculate the finite-time asymptotic Landau-Zener probability. Therefore, we establish a finite-time error regime depicted in the main panel of Figs.~\ref{fig4}(a-b) as error bars using the experimental values of the annealing time. The estimation of the Landau-Zener jump-time $t^{\star}$ has been implemented uniquely from the theoretical prediction, assuming it will be the same for the experimental data. Note that the adiabatic-impulse approximation and the asymptotic Landau-Zener probability are equivalent in the regime of our experimental data giving confidence in our validation of the KZM on the IBM-Q platform.\\
\\
\begin{figure}[h!]
\centering
\includegraphics[width=0.6\linewidth]{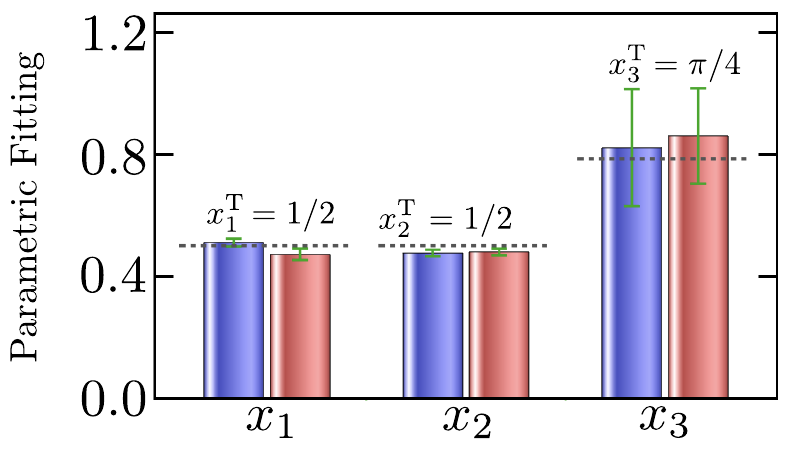}
\caption[justification=justified]{\label{fig7} {\bf KZM adiabatic-impulse approximation fitting parameters.}  From the best qubit experimental data (solid symbols in Fig.~\ref{fig4}), the fitting to the KZM adiabatic-impulse approximation $P_{AI}(t_a)$ given by Eq.\eqref{fit_M} is probed (the dashed gray lines correspond to the theoretical predictions). The experimental data at \texttt{ibmq\_bogota} and \texttt{ibmq\_lima} are depicted in blue and red, respectively. }
\end{figure}

 For the qubit with the largest decoherence $T_1$ and $T_2$ times (the best qubit from now on), the experimental data show an excellent agreement with the theoretical predictions for the impulse-adiabatic approximation. For large annealing time $t_a$, the experimental data has a significant deviation  for some qubits in the \texttt{ibmq\_bogota} quantum computer. Indeed, the adiabatic-impulse approximation relationship with the Landau-Zener problem assumes a close system's quantum dynamics. However, since IBM-Q is benchmarked as an open-quantum system, deviations are to be expected.\\

 In order to further testing the KZM adiabatic-impulse approximation, from our experimental data, we rewrite the Eq.~\eqref{PAI_1} in terms of $3$ fitting parameters, as
\begin{equation}\label{fit_M}
P_{AI}(t_a) \simeq x_1-x_2\sqrt{1-\frac{2}{\pap{x_3 t_a}^2+x_3 t_a\sqrt{\pap{x_3 t_a}^2 +4}+2}}.
\end{equation}
In Fig.~\eqref{fig7}, we depict the comparison of the fitting parameters $x_1$, $x_2$ and $x_3$ for the best qubit at \texttt{ibmq\_bogota} and \texttt{ibmq\_lima}. The structure of the fitting expression allows us a direct comparison with the theoretical predictions $\pap{x_1^{{\rm T}}, x_2^{{\rm T}}, x_3^{{\rm T}}}$. The first fitting parameter $x_1$ provides information about how robust the qubit is to decoherence for fast LZ driving. Note that the theoretical prediction is $x_1^{{\rm T}}=1/2$ as it is fixed by the initial condition at the anticrossing initial point. Moreover, it fixes the value of the adiabatic-impulse approximation for small annealing times, $P_{{\rm AI}}\pap{t_a\to0}=1/2$. It is evident from Fig.~\ref{fig4} that some qubits deviate from this ideal value in this regime, confirming that these qubits are already highly sensible to decoherence. Nonetheless, for these results, we used the smallest number of layers considered. The second fitting parameter $x_2$ gives information about the higher annealing time regime, with theoretical value $x_2^{{\rm T}} = 1/2$. The asymptotic value of the adiabatic-impulse approximation is zero for large annealing times. However, large annealing times imply that the LZ transition probability has several oscillations as a function of time. Consequently, it is necessary to manage large simulation times to obtain the asymptotic LZ probability. It is to be expected that, our results show deviations due to finite simulation time effects. Finally, the third parameter $x_3$ validates the Kibble-Zurek scaling in the adiabatic-impulse approximation ($x_3^{{\rm T}} = \pi/4$).  We found an excellent agreement with the theoretical predictions for these $\mathcal{QC}_1$ and $\mathcal{QC}_2$ robust qubits.  Thus, by using the close relationship between the KZM and the LZ transition probability, we validated and tested the KZM on IBM-Q. These results can be part of a sequence of major steps to fully understand the strength and limitations of time-dependent quantum simulations. It may provide insights for designing top efficient quantum simulation protocols for more involved out-of-equilibrium and interacting systems.

\section{Conclusions}\label{Conclu}
In this work we explored the dynamics of a two level system under the time-dependent Landau-Zener Hamiltonian on digital IBM Quantum computers. Time evolution was simulated by discretization of the time dependent Hamiltonian and application of subsequent single-qubit unitary gates representing finite time progressions. We studied the Landau Zener transition probability as a function of time by running parallel quantum circuits on 5-qubit machines \texttt{ibmq\_lima} and \texttt{ibmq\_bogota} with different topologies. We find a strong agreement with the theoretical solution of the LZ problem for robust qubits from both machines. We also considered the effect of decoherence on an open LZ system, modeling the dissipation using collapse operators for relaxation and dephasing. For greater trotterizations of the time evolution operator, increasing computing time cause noticeable deviations from the theoretical LZ solution. The numerical solution of the Lindblad master equation accurately depicts the open system's relaxation towards the ground state, supported by the measured probabilities.\\
\\
The above positive LZ results allowed us to demonstrate the first simulation on a realistic quantum computer of the universal Kibble-Zurek mechanism by estimating the asymptotic transition probability obtained from LZ experimental data. Results show excellent agreement for the best qubits considered in each device and low annealing times. We find that larger annealing times demand a greater time resolution in the evolution operator discretization, putting practical limits on the performance achieved, as it becomes limited by the conflict between computing depth and decoherence times. However, the rapid rate of quantum hardware advances may soon change this. Furthermore, an interesting follow-up research direction would consist in focusing on richer open quantum platforms, where KZM has been poorly explored. Thus, using real quantum hardware to test quantum universal dynamical behaviors, in both closed and open systems, represent an interesting extension of the results presented in this work.
\section*{Data Availability Statement}
The datasets presented in this study can be found in online repositories. The names of the repository/repositories and accession number(s) can be found below: \url{https://github.com/sanhq17/Testing_KZM_IBMQ}.
\section*{Author Contributions}
FG-R and LQ initiated and guided the project. SH-Q took the experimental measurements. FG-R developed numerical simulations and prepared the figures. All authors contributed to the analysis of the results and the writing of the manuscript.
\section*{Funding}
S.H-Q, F.J.R. and L.Q. are thankful for the financial support from Facultad de Ciencias-UniAndes projects: INV-2021-128-2292, and INV-2019-84-1841. F.J.G-R acknowledges financial support from European Commission FET-Open project AVaQus GA 899561.
\section*{Acknowledgments}
The authors thank to Bogdan Damski for useful comments and suggestions.
\section*{Conflict of Interest Statement}
The authors declare that the research was conducted in the absence of any commercial or financial relationships that could be construed as a potential conflict of interest.
\section*{Publisher’s note}
All claims expressed in this article are solely those of the authors and do not necessarily represent those of their affiliated organizations, or those of the publisher, the editors and the reviewers. Any product that may be evaluated in this article, or claim that may be made by its manufacturer, is not guaranteed or endorsed by the publisher.
\section*{Supplemental Data}
The Supplementary Material for this article can be found online at: \url{https://www.frontiersin.org/articles/10.3389/frqst.2022.1026025/full#supplementary-material}

\bibliography{My_bib_IBM_LZ}{}
\bibliographystyle{apsrev4-1}
\newpage
\pagebreak
\clearpage
\setcounter{equation}{0}
\setcounter{figure}{0}
\setcounter{table}{0}
\setcounter{section}{0}
\setcounter{page}{1}
\makeatletter
\renewcommand{\thefigure}{\arabic{figure}}
\renewcommand{\figurename}{{\bf Supplementary Figure.}}
\renewcommand{\theequation}{S\arabic{equation}}
\renewcommand{\thesection}{{\large {\bf Supplemental Material \arabic{section}}}}
\renewcommand{\thesubsection}{\arabic{subsection}}
\renewcommand{\bibsection}{\section*{Supplementary Material: References}}

\newcounter{fnnumber}
\renewcommand{\thefootnote}{\fnsymbol{footnote}}
\begin{center}
\textbf{\large ---Supplementary Material---\\
Experimental validation of the Kibble-Zurek Mechanism on a Digital Quantum Computer
}\\
\vspace{0.5cm}
Santiago Higuera-Quintero\href{https://orcid.org/0000-0001-6905-4662}{\includegraphics[scale=0.45]{orcid}}\,$^{1}$, Ferney J. Rodr\'iguez\href{https://orcid.org/0000-0001-5383-4218}{\includegraphics[scale=0.45]{orcid}}\,$^{1}$, \\
Luis Quiroga\href{https://orcid.org/0000-0003-2235-3344}{\includegraphics[scale=0.45]{orcid}}\,$^{1}$, and Fernando J. G\'omez-Ruiz\href{https://orcid.org/0000-0002-1855-0671}{\includegraphics[scale=0.45]{orcid}}\,$^{2,*}$\\
\vspace{0.2cm}
$^{1}${\it Departamento de F{\'i}sica, Universidad de los Andes, A.A. 4976, Bogot\'a D. C., Colombia}\\
$^{2}${\it Instituto de F\'isica Fundamental IFF-CSIC, Calle Serrano 113b, Madrid 28006, Spain}\\
$^{*}${\it Corresponding Author:} \href{mailto:fernandojavier.gomez@iff.csic.es}{fernandojavier.gomez@iff.csic.es}
\end{center}
\section{Landau-Zener Formal Solution}\label{Append1}
Here, we present a way to calculate the LZSM transition probability depending on the initial conditions. Consider the time-dependent Schr\"odinger equation
\begin{equation}\label{Sch_EQ}
i\frac{d}{dt}\ket{\psi\pap{t}}=\hat{H}\pap{t}\ket{\psi\pap{t}}.
\end{equation}
Where, the Hamiltonian $\hat{H}\pap{t}$ is given by: 
\begin{equation}\label{LZ_Hamil}
\hat{H}\pap{t}=-\frac{\varepsilon\pap{t}}{2}\hat{\sigma}_{z}-\frac{\Delta}{2}\hat{\sigma}_x.
\end{equation}
We write the wave function as a linear combination of diabatic states given by
\begin{equation}\label{Wav_LZ}
\ket{\psi\pap{t}}=\alpha\pap{t}\ket{0}+\beta\pap{t}\ket{1}.
\end{equation}
In general, we consider the time-evolution from an initial time $t_i$ to a final time $t_f$. Therefore, the wave-function initial condition is fixed by  $\ket{\psi\pap{t_i}} =\alpha_i \ket{0} + \beta_i \ket{1}$. The probability amplitudes $\alpha_i$ and $\beta_i$ satisfies that $\left|\alpha_i\right|^2+\left|\beta_i\right|^2=1$. By direct substitution of Eq.~\eqref{Wav_LZ} into Eq.~\eqref{Sch_EQ}, we obtain the system of differential equations
\begin{subequations}
\begin{align}
i\frac{d}{dt}\alpha\pap{t}&= -\frac{\varepsilon\pap{t}}{2}\alpha\pap{t} -\frac{\Delta}{2}\beta\pap{t},\label{alphat}\\
i\frac{d}{dt}\beta\pap{t}&= -\frac{\Delta}{2}\alpha\pap{t} +\frac{\varepsilon\pap{t}}{2}\beta\pap{t}.\label{betat}
\end{align}
\end{subequations}
Decoupling the differential equations, we obtain
\begin{equation}
\begin{split}
\frac{d^2\alpha\pap{t}}{dt^2}&= -\pas{\pap{\frac{\varepsilon\pap{t}}{2}}^2 +\pap{\frac{\Delta}{2}}^2 +\frac{i}{2}\frac{d\varepsilon\pap{t}}{dt}}\alpha\pap{t},\\
\frac{d^2\beta\pap{t}}{dt^2}&= -\pas{\pap{\frac{\varepsilon\pap{t}}{2}}^2 +\pap{\frac{\Delta}{2}}^2 -\frac{i}{2}\frac{d\varepsilon\pap{t}}{dt}}\beta\pap{t}.
\end{split}
\end{equation}
Using the linear bias $\varepsilon\pap{t}$ dependence and the substitution $t=\sqrt{2t_a}\tau$, we rewrote the previous differential equations in the form of two parabolic cylinder differential equation:
\begin{subequations}
\begin{align}
\frac{d^2}{d\tau^2}\alpha\pap{\tau}+\pap{2\delta + i +\tau^2}\alpha\pap{\tau}&=0.\label{Eq_Weber1}\\
\frac{d^2}{d\tau^2}\beta\pap{\tau}+\pap{2\delta - i +\tau^2}\beta\pap{\tau}&=0,\label{Eq_Weber2}
\end{align}
\end{subequations}
Where, we defined $\delta = \Delta^2 t_a/4$ as the adiabaticity parameter. The canonical form of the parabolic cylinder differential equation is the second-order ordinary differential equation
\begin{equation}\label{st_PCD}
\frac{d^2}{dz^2}u\pap{z}+\pap{p+\frac{1}{2}-\frac{z^2}{4}}u\pap{z}=0,
\end{equation}
whose solution is given by
\begin{equation}
u\pap{z}=c_1 \mathbf{D}_p\pap{z} + c_2 \mathbf{D}_{-p-1}\pap{iz},
\end{equation}
where $ \mathbf{D}_p\pap{z}$ is a parabolic cylinder function and the constants $c_1 \pap{c_2}$ depend on the initial conditions~\cite{Weber_69}. Additionally, The Weber's equation (Eq.~\eqref{st_PCD}) has a symmetry by simultaneously replace $p$ and $z$ by $-p-1$ and $\pm i z$ respectively~\cite{whittaker21}. Therefore, $\mathbf{D}_p\pap{-z}$, $\mathbf{D}_{-p-1}\pap{-iz}$ are solutions of Weber's equation. These four solutions are linearly dependent~\cite{Gradshteyn}. We transform the Eq.\eqref{Eq_Weber1} and Eq.\eqref{Eq_Weber2} in the form of Eq.~\eqref{st_PCD} using the substitution $z=\sqrt{2}\tau\exp\pas{i\pi/4}$. We obtained
\begin{subequations}
\begin{align}
\frac{d^2}{dz^2}\alpha\pap{z}+\pap{-i\delta - \frac{1}{2} -\frac{z^2}{4}}\alpha\pap{z}&=0\label{Eq_Weber1A},\\
\frac{d^2}{dz^2}\beta\pap{z}+\pap{-i\delta + \frac{1}{2} -\frac{z^2}{4}}\beta\pap{z}&=0\label{Eq_Weber2A}.
\end{align}
\end{subequations}
We propose a formal solution of Eq.~\eqref{Eq_Weber2A} given by
\begin{equation}\label{sol_beta}
\beta\pap{z}= \chi_1 \mathbf{D}_{-i\delta}\pap{z} + \chi_2 \mathbf{D}_{-1+i\delta}\pap{iz}.
\end{equation}
Where, $\chi_1$ and $\chi_1$ are the initial conditions at $z=z_i$. On the other hand, we rewrote Eq.~\eqref{betat} as
\begin{equation}\label{alpha2}
\alpha\pap{z}=\frac{e^{-i\frac{\pi}{4}}}{\sqrt{\delta}}\pas{\frac{d}{dz}\beta\pap{z} + \frac{1}{2}z\beta\pap{z}}.
\end{equation}
By direct substitution of Eq.~\eqref{sol_beta} into Eq.~\eqref{alpha2}, and using the recurrence formulae given by
\begin{subequations}
\begin{align}
\mathbf{D}_{n+1}\pap{z}-z \mathbf{D}_{n}\pap{z}+n\mathbf{D}_{n-1}\pap{z} &=0\label{Eq_recurrence1}\\
\mathbf{D}_{n}'\pap{z}+\frac{1}{2}z\mathbf{D}_{n}\pap{z}-n\mathbf{D}_{n-1}\pap{z}&=0,\label{Eq_recurrence2}
\end{align}
\end{subequations}
we obtain
\begin{equation}\label{Eq:expressionAlpha}
\alpha\pap{z}=\frac{e^{-i\frac{3\pi}{4}}}{\sqrt{\delta}}\left[\delta\chi_1\mathbf{D}_{-1-i\delta}(z) + \chi_2\mathbf{D}_{i\delta}(iz)\right].
\end{equation}
Now, we can find $\chi_1$ and $\chi_2$ from the initial conditions at $z=z_i$ with
certain $\alpha\pap{z_i}$ and $\beta\pap{z_i}$:
\begin{align}
\chi_1 &= \frac{e^{i\frac{3\pi}{4}}\sqrt{\delta} \mathbf{D}_{-1+i\delta}\pap{i z_i}\alpha\pap{z_i}-\mathbf{D}_{i\delta}\pap{i z_i}\beta\pap{z_i}}{\delta \mathbf{D}_{-1-i\delta}\pap{z_i}\mathbf{D}_{-1+i\delta}\pap{i z_i} -\mathbf{D}_{-i\delta}\pap{z_i} \mathbf{D}_{i\delta}\pap{i z_i}},\label{coeff_b1}\\
\chi_2 &= \frac{-e^{i\frac{3\pi}{4}}\sqrt{\delta}\mathbf{D}_{-i\delta}(z_i) \alpha\pap{z_i} + \delta\mathbf{D}_{-1-i\delta}(z_i)\beta(z_i)}{\delta \mathbf{D}_{-1-i\delta}\pap{z_i}\mathbf{D}_{-1+i\delta}\pap{i z_i} -\mathbf{D}_{-i\delta}\pap{z_i} \mathbf{D}_{i\delta}\pap{i z_i}}.\label{coeff_b2}
\end{align}

\subsection{Exact solution of LZ evolution starting at the anticrossing state}\label{SolutionAnti}
We are interested in the evolution when the system is in the ground state at $t=0$. The state is given by  $\ket{\psi\pap{0}}=\pap{\ket{0}-\ket{1}}/\sqrt{2}$. Therefore, $\alpha\pap{z_i}=1/\sqrt{2}$ and $\beta\pap{z_i}=-1/\sqrt{2}$. Additionally, we can express $z$ as function of $t$ as $z=\frac{t}{\sqrt{t_a}}\exp\pas{i\pi/4}$. Therefore, we evaluate the following limits
\begin{equation}
\lim_{z_i\to 0} \mathbf{D}_n \pap{\pm z_i}=\lim_{z_i\to 0} \mathbf{D}_n \pap{\pm i z_i}=\frac{2^{n/2}\sqrt{\pi}}{\Gamma\pap{\frac{1-n}{2}}},\label{limit1}
\end{equation}
\begin{equation}
 \lim_{z_i\to0}\pas{\delta \mathbf{D}_{-1-i\delta}\pap{z_i}\mathbf{D}_{-1+i\delta}\pap{i z_i} -\mathbf{D}_{-i\delta}\pap{z_i} \mathbf{D}_{i\delta}\pap{i z_i}}=-\exp\pas{-i\pi k},\quad\text{with}\quad k=-\frac{i\delta}{2}.
\end{equation}

Therefore, we obtain the values of $\chi_1$ and $\chi_2$ as:
\begin{align*}
\chi_1 &= -\frac{2^{k}\exp\pas{i\pi k}}{4\sqrt{ik}}\pas{\frac{\sqrt{2ik}\Gamma\pap{k}+\pap{1+i}\Gamma\pap{\frac{1}{2}+k}}{\Gamma\pap{2k}}},\\
\chi_2&= \frac{\exp\pas{i\pi k}}{2^{k+1}}\pas{\frac{2ik\Gamma\pap{\frac{1}{2}-k}+\pap{1-i}\sqrt{2ik}\Gamma\pap{1-k}}{\Gamma\pap{1-2k}}}.
\end{align*}
Here,  we used the following Gamma-function relations
\begin{align}
2^{2z-1}\Gamma\pap{z}\Gamma\pap{\frac{1}{2}+z}&=\sqrt{\pi}\Gamma\pap{2z},\\
\Gamma\pap{1-z}&=-z\Gamma\pap{-z}.
\end{align}
Now we calculate the asymptotic transition probability to the higher energy eigenstate
\begin{equation}\label{Eq:transitionProbability}
    \left|\left\langle1\big\rvert\psi\left({t\to\infty}\right)\right\rangle\right|^2 = 1 - \left|\alpha\left({t\to\infty}\right)\right|^2,
\end{equation}
using the expansions for parabolic cylinder functions~\cite{Gradshteyn}

\begin{align}
\mathbf{D}_p(z) &\sim e^{-z^2/4}z^p-\frac{\sqrt{2\pi}}{\Gamma\pap{-p}}e^{i\pi p}e^{z^2/4}z^{-p-1}\quad\text{for}\quad\frac{\pi}{4}<\arg\pap{z}<\frac{5\pi}{4},\label{limit_weber_1}\\
\mathbf{D}_p(z)&\sim e^{-z^2/4}z^p\quad\text{for}\quad\left|\arg\pap{z}\right|<\frac{3\pi}{4}.\label{limit_weber_2}
\end{align}

whenever $|z|\to\infty$. We obtain the limiting values for the relevant parabolic cylinder functions
\begin{align}
    \mathbf{D}_{-1-i\delta}\pap{z\big\rvert_{t\to\infty}} &= 0,\\
    \mathbf{D}_{i\delta}\pap{iz\big\rvert_{t\to\infty}} &= e^{-\frac{3\pi\delta}{4} + i\phi(t)},
\end{align}
where $\phi(t) = t^2/4t_a + \delta\log\pap{t/\sqrt{t_a}}$ is a time dependent phase. By substitution into Eq.~\eqref{Eq:expressionAlpha} and \eqref{Eq:transitionProbability}, we obtain the asymptotic transition probability
\begin{equation}
    P_{LZ}(t\to\infty) = 1-\frac{e^{-\frac{3\pi\delta}{2}}}{\delta}|\chi_2|^2.
\end{equation}
\subsection{Exact solution of LZ evolution starting from the ground state}\label{SolutionTotal}
In the classical LZ problem, evolution starts from the ground state $\ket{0}$ of the LZ Hamiltonian at $t= -\infty$. Then, we may set $\alpha(-\infty) = 1$ and $\beta(-\infty) = 0$ as initial conditions. Using the asymptotic expansions Eq.~\eqref{limit_weber_1} and \eqref{limit_weber_2} and substitution into relations \eqref{coeff_b1} and \eqref{coeff_b2}, we determine the coefficients corresponding to the classical LZ problem initial conditions
\begin{align}
    \chi_1 &= 0,\\
    \chi_2 &= \sqrt{\delta}e^{-\pi\delta/4}.
\end{align}
Given this result, the transition probabilities can be written compactly as
\begin{align}
    \left|\langle0|\psi(z)\rangle\right|^2 &= e^{-\pi\delta/2}\left|\mathbf{D}_{i\delta}\pap{iz}\right|^2,\label{probability0}\\
    \left|\langle1|\psi(z)\rangle\right|^2 &= \delta e^{-\pi\delta/2} \left|\mathbf{D}_{-1+i\delta}\pap{iz}\right|^2,\label{probability1}
\end{align}
and the classical LZ formula can directly recovered by applying the asymptotic expansion \eqref{limit_weber_2} on Eq.~\eqref{probability0}, where we obtain
\begin{equation}
    P_\text{LZ} = e^{-2\pi\delta}.
\end{equation}

\section{Systematic Readout Error Mitigation}\label{sec:ReadoutErrorMitigation}
Results obtained from the available quantum hardware are subject to multiple sources of error including thermal relaxation, gate errors and faulty readout of the prepared quantum state \cite{mitigationBultrini}. A common first order approach to mitigate systematic readout errors is through a qubit's calibration matrix. For a single qubit this is defined to be
\begin{equation}
    \mathbf{A} = \begin{pmatrix}
    p_{00} & p_{01}\\
    p_{10} & p_{11}
    \end{pmatrix},
\end{equation}
where $p_{ij}$ are the probabilities that a qubit prepared in state $\ket{j}$ is measured in state
$\ket{i}$, for an ideal quantum computer this would be equal to the identity matrix. For a particular qubit of an IBM Quantum Computer, the calibration matrix can be found from the computer's system properties. Properties \texttt{prob\_meas1\_prep0} and \texttt{prob\_meas0\_prep1} correspond to $p_{10}$ and $p_{01}$ respectively. The diagonal terms of the calibration matrix can then be determined as the sum of elements in the columns must be equal to 1.

Systematic errors for a great number of executions can then be mitigated by inversion of the calibration matrix since $\Vec{P}_\text{noisy} = \mathbf{A}\Vec{P}_\text{ideal}$. Thus a useful calibration formula to infer the ideal results is
\begin{equation}
    \Vec{P}_\text{ideal} = \mathbf{A}^{-1}\Vec{P}_\text{noisy},
\end{equation}
where $\Vec{P}_\text{noisy} = [P_0,P_1]^\top$ is the vector of experimental probabilities $P_i$ of the measured quantum state.

\end{document}